\documentclass[prd,aps,twocolumn,reprint,floatfix,nofootinbib,superscriptaddress]{revtex4-1}
\usepackage{amssymb,graphicx}
\usepackage{epsfig}
\usepackage[usenames]{color}

\newcommand{\beqn}{\begin{eqnarray}}
\newcommand{\eeqn}{\end{eqnarray}}
\newcommand{\beq}{\begin{equation}}
\newcommand{\eeq}{\end{equation}}

\begin{document}

\title{Black hole superradiance in dynamical spacetime}
\author{William E.\ East}	
\email{weast@stanford.edu}
\affiliation{
Kavli Institute for Particle Astrophysics and Cosmology, Stanford University, SLAC National Accelerator Laboratory, Menlo Park, California 94025, USA
}
\author{Fethi M.\ Ramazano\u{g}lu}
\affiliation{Department of Physics, Princeton University, Princeton, New Jersey 08544, USA}
\author{Frans Pretorius}
\affiliation{Department of Physics, Princeton University, Princeton, New Jersey 08544, USA}

\begin{abstract}
We study the superradiant scattering of gravitational waves by a nearly
extremal black hole (dimensionless spin $a=0.99$) by numerically solving
the full Einstein field equations, thus including backreaction effects.
This allows us to study the dynamics of the black hole as it loses energy and angular momentum
during the scattering process. To explore the nonlinear phase of the interaction,
we consider gravitational wave packets with initial energies up to $10\%$ of the mass of 
the black hole. We find that as the incident wave energy increases, the 
amplification of the scattered waves, as well as the energy extraction efficiency 
from the black hole, is reduced. During the interaction the apparent horizon
geometry undergoes sizable nonaxisymmetric oscillations. The largest
amplitude excitations occur when the peak frequency of the incident wave packet is
above where superradiance occurs, but close to the dominant quasinormal mode frequency of the black hole.
\end{abstract}

\maketitle

\section{Introduction}
Despite the fact that spinning black holes (BHs) serve as one-way membranes to causal curves, 
it is possible to extract rotational energy from them.
This can occur through the Penrose process~\cite{Penrose69} for particles, 
or by scattering waves off a BH in an analogous process called 
superradiance~\cite{PhysRevLett.28.994,1973ZhETF..64...48S,1972JETP...35.1085Z}.
Superradiance occurs for scalar fields, 
electromagnetic fields, and gravitational waves (GWs), the latter
having the strongest possible amplification~\cite{Teukolsky:1974yv}.
This idea has implications across numerous fields from astrophysics --- 
where this energy extraction has been invoked 
to explain the powering of astrophysical jets 
in the Blandford-Znajek effect and related processes~\cite{Blandford:1977ds, Ruffini:1975ne} --- to quantum gravity, high-energy physics, and the AdS/CFT correspondence~\cite{DeWitt1975295,PhysRevD.61.024014,Bredberg:2010}.

Superradiance can occur in nongravitational settings as well (e.g. the interaction
of electromagnetic radiation with a rotating boundary), and it can be
argued that amplification in certain regimes is required
by the second law of thermodynamics~\cite{Bekenstein:1998nt}.
For BHs, the analogous argument derives from the connection between
horizon area and entropy, and Hawking's area increase theorem~\cite{PhysRevD.7.949}.
At the linear level, a wave with frequency $\omega$, azimuthal 
number $m$, and incident energy $\delta E$ will have 
angular momentum $\delta J = m \delta E / \omega$.  
The change in a BH's area for a given change in mass and angular momentum 
is given by $dA=(8\pi/\kappa)(dM - \Omega_{\rm BH} dJ)$,where 
$\Omega_{\rm BH}$ is the rotational frequency of the BH and $\kappa$ is its surface gravity
(we use geometric units with $G=c=1$ throughout). Thus if 
a wave with frequency $0<\omega < m \Omega_{\rm BH}$ were wholly absorbed 
the BH's horizon area would decrease, in violation of the
second law. What happens instead is
the wave is scattered, gaining energy and angular momentum,
and the BH effectively absorbs what globally is
counted as negative energy and angular momentum (as also
occurs in the Penrose process).

Despite the long history of study of this phenomenon in general relativity,
essentially no work has delved beyond the linear level to self-consistently 
include backreaction, and address how the properties of the spacetime change to balance
the energy and angular momentum carried off by the scattered waves.
There are several reasons why this 
is of interest. To begin with, for highly spinning BHs, waves in, and slightly above
the superradiant frequency range, carry a ratio of energy to angular momentum such that 
if they were absorbed they could overspin the BH~\cite{Duztas:2013wua}.  
Hence the role of backreaction is important
to the question of cosmic censorship.  Additionally, 
superradiance is related to the hypothesized existence of so-called floating orbits~\cite{1972Natur.238..211P},
originally suggested by Misner, 
where an object orbiting a spinning BH extracts energy from the BH at the same
rate it emits energy to infinity via GWs.
Recent work suggests that these orbits do not exist at least
in the extreme-mass-ratio case~\cite{PhysRevD.87.044050} (though they may 
exist for matter coupled to a massive scalar field~\cite{2011PhRvL.107x1101C}). 
However, since very little is known about superradiance
in the context of a dynamical spacetime, floating orbits cannot yet be ruled out 
in general.
Finally, superradiance is related to the idea of a black hole ``bomb," in which a spinning
BH is enclosed in a reflective cavity and perturbed with a superradiant mode,
which will grow upon each successive reflection until a sizable fraction of the rotational
energy has been transferred to the waves~\cite{1972JETP...35.1085Z,1972Natur.238..211P}
(though cf.~\cite{Witek:2010qc}).  
A similar effect occurs with light bosonic fields, where self-interaction due to
the mass term plays the role of the 
confiner~\cite{Damour:1976,Detweiler:1980uk,Zouros:1979iw,Witek:2012tr}.
If such fields exists, rapidly rotating astrophysical BHs with radii commensurate
with the Compton wavelength of the field would be superradiantly unstable,
and backreaction 
would become important to the late-time dynamics.

Of course, since a BH has a limited amount of rotational energy that can be
extracted without decreasing the BH's area, it is expected that the efficiency
of superradiant amplification will decrease as the amplitude of the incident wave
increases. Also, the above thermodynamic arguments used the laws of BH
mechanics for quasistatic processes, hence it is especially interesting to 
explore the behavior of superradiance in the regime where the BH is highly dynamical.
In this paper, we study the superradiant scattering of a gravitational
wave packet by a highly spinning BH using numerical solutions of 
the full Einstein equations. For low amplitude waves, we find results consistent
with linear theory, with an amplification (for the chosen parameters)
of $\approx 40\%$. As the amplitude increases, we find 
that both the amplification and rotational energy extraction efficiency
decrease. During the interaction, the apparent horizon (AH)
geometry develops high-amplitude, nonaxisymmetric oscillations,
the largest of which occur 
when the peak frequency of the wave packet is near the least-damped
quasinormal (resonant) mode frequency of the BH.

\section{Methodology}
We solve the field equations in the generalized harmonic formulation, 
using the code described in~\cite{Pretorius:2004jg, code_paper}.
As a gauge choice, we fix the source functions $H^a = \Box(x^a)$ to be those of the 
isolated BH.
The numerical grid we use has six refinement levels with a grid spacing of $\approx 0.03M$ on the finest
level and $\approx 2.4M$ in the wave zone, where $M$ is the initial mass of the BH. 
For select cases, we also run simulations with 0.6 and 0.8 times the resolution to 
establish convergence. 

We construct initial data representing a BH with an incoming GW packet.  
We use a BH in Kerr-Schild coordinates with dimensionless spin $a=0.99$. 
To determine the initial metric perturbation to the BH spacetime representing the GW,
we begin with an ingoing, spin-2 solution to the Teukolsky equation which is characterized by 
spheroidal harmonic indices $l,m$, and the frequency $\omega$~\cite{Teukolsky:1974yv},
the angular part of which we numerically calculate as in~\cite{2006PhRvD..73b4013B,1985RSPSA.402..285L}.   
For the radial part, we use the approximate asymptotic form~\cite{2003PhRvD..67l4010O},
which introduces negligible error for a wave packet peaked far
from the BH.  
This gives the value for the Newman-Penrose scalar $\Psi_0$, which we then convert
into a metric perturbation using the techniques of~\cite{2002PhRvD..66b4026L,2003PhRvD..67l4010O}.
We multiply this by a Gaussian envelope in the radial direction 
peaked at $r_0/M=40$ with standard deviation $\sigma/M=10$. 
To avoid regularity problems at the BH horizon
when transforming the perturbation to Kerr-Schild coordinates, we also multiply by 
a function that smoothly goes from one to zero at a radius of $\approx 3M$ from the origin.
Finally, we use this solution to specify free data for the constraint equations
in the conformal thin-sandwich formulation, which we solve as in~\cite{idsolve_paper}.

We focus on $l=2$ wave packets and in order to maintain equatorial symmetry, we use the superposition of $(m,\omega)=(\pm2,\pm\omega_0)$ solutions.
This $l$ and $m$,  along with our choice of $\omega_0 M = 0.75$, is chosen to give near maximal amplification
given our frequency spread.
We consider wave packets with a range of amplitudes.
The lowest amplitude wave has a mass of $\approx  5.8 \times 10^{-4}M$.  Normalizing to this lowest value, the other waves have
amplitudes $A= 2, 4, 8$, and $12$ times this value.  The mass of a wave packet scales as $A^2$ and the
largest amplitude case has a mass of $0.098M$.
We perform resolution studies for the $A=4$ and $A=12$ cases.
We also perform simulations with $\omega_0 M= 0.865$ and $\omega_0 M=1$
(near and above the maximum superradiant frequency), and amplitudes chosen to give
the same mass as the $A=12$ case.  We measure the outgoing GW radiation resulting from the interaction 
of the wave packet with the BH 
by calculating the Newman-Penrose scalar $\Psi_4$ (see e.g.~\cite{Smarr79})
and monitor the BH AH during the scattering process. 
 
\section{Results}
\begin{figure}
\begin{center}
\includegraphics[width=\columnwidth,draft=false]{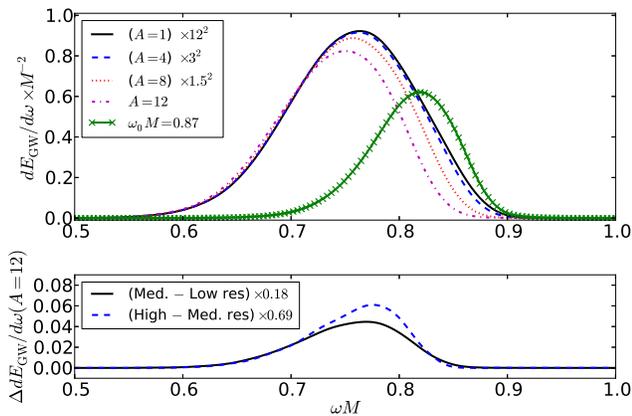}
\end{center}
\caption{
Top: Energy spectrum of outgoing GWs after the scattering of an ingoing GW 
packet with central frequency $\omega_0 M=0.75$
and various amplitudes.  The lower amplitude waves have been rescaled according
to their leading-order dependence on ingoing amplitude.  We also show a case
with $\omega_0 M=0.87$ and the same initial energy as the $A=12$ case.
Bottom: Truncation error for the $A=12$ case, consistent with between third- and fourth-order
convergence (the error has been scaled assuming the latter).
\label{eomega_fig}
}
\end{figure}

For the smaller amplitude cases ($A\lesssim4$) with $\omega_0 M=0.75$, the energy and angular momentum of
the outgoing GWs are approximately proportional to the ingoing values (i.e., they scale with $A^2$) 
as expected in linear perturbation theory.
Using the difference between the BH horizon mass and the ADM mass as a measure
of the initial mass of the GW wave packet, we estimate in this regime (in particular for $A=4$) the 
outgoing GW carries an extra $44 \pm 7 \%$  
more energy than the ingoing GW.
This is consistent with the $\approx 40\%$ expected from linear perturbation theory~\cite{Teukolsky:1974yv}
for such an energy spectrum.
However, as seen in Fig.~\ref{eomega_fig}, for larger amplitude cases the scaling begins to break
down and there is less energy at higher frequencies relative to the low amplitude cases.
This trend is consistent with the fact that (as we discuss below) the spin of the BH is noticeably decreasing 
for these cases and BHs with lower spins are less superradiant at higher frequencies.

\begin{figure}
\begin{center}
\includegraphics[width=\columnwidth,draft=false]{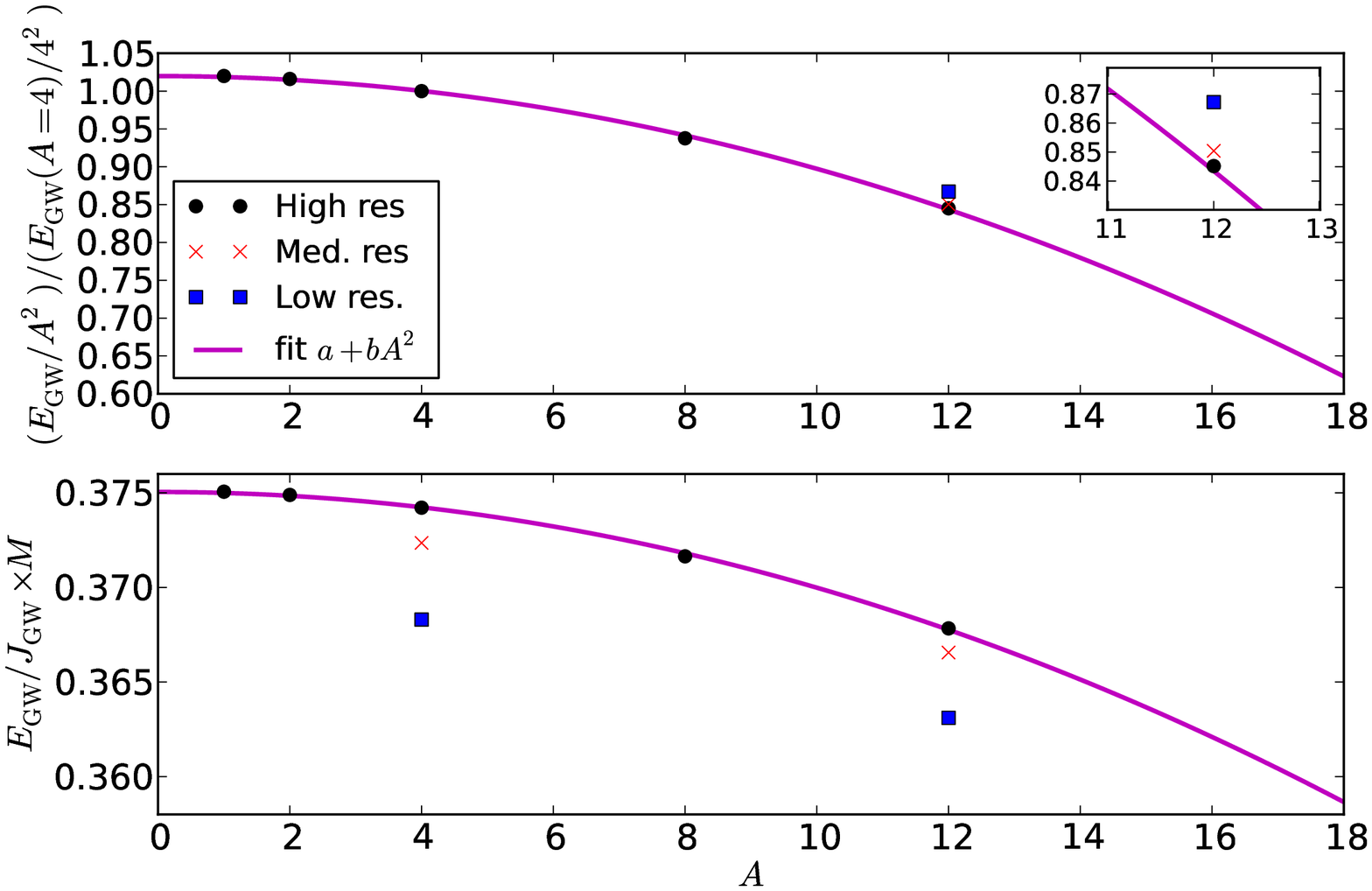}
\includegraphics[width=\columnwidth,draft=false]{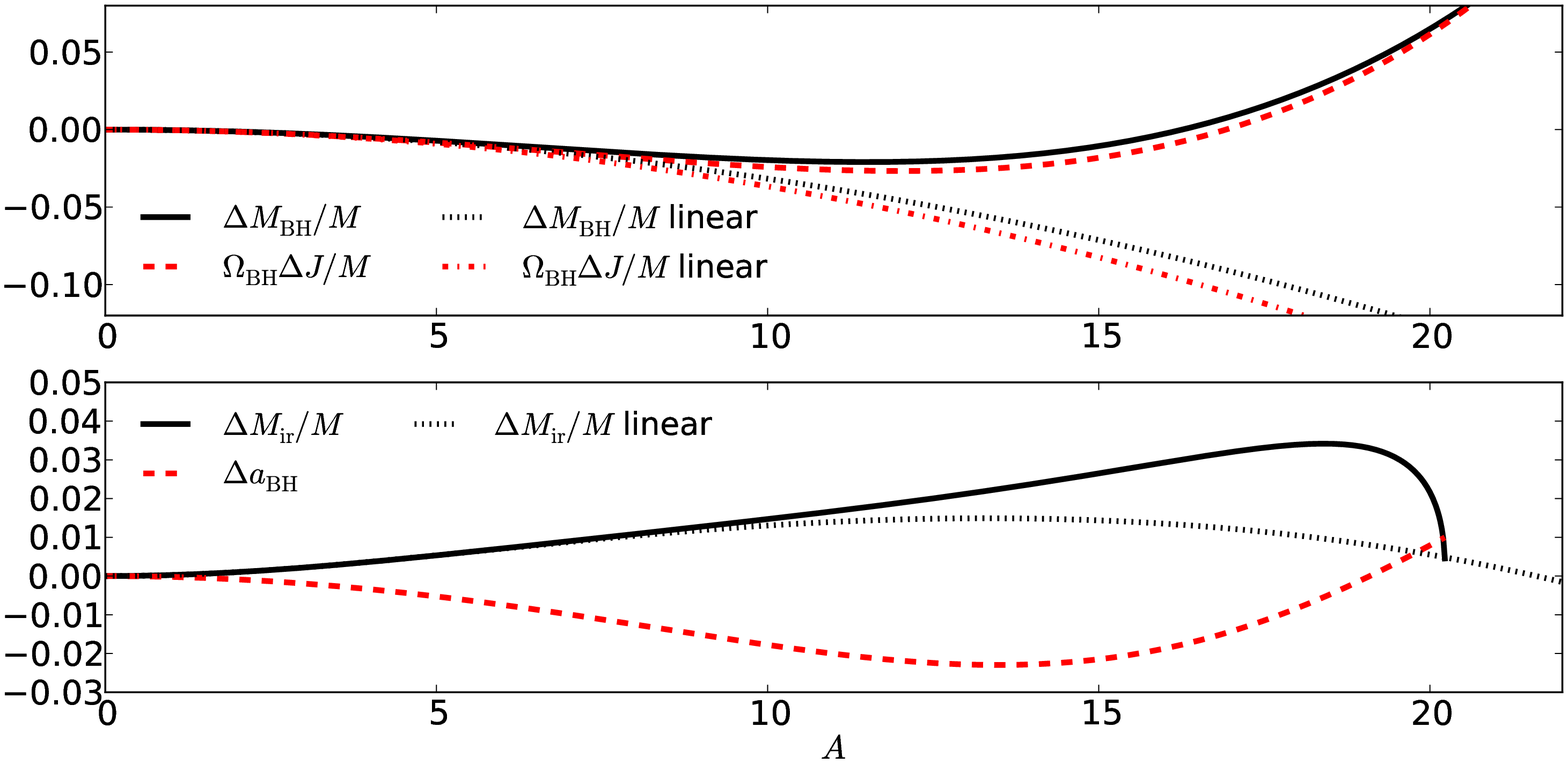}
\end{center}
\caption{
Top: The energy in outgoing GWs for different values of the ingoing amplitude $A$, 
relative to the value obtained by scaling the $E_{\rm GW}$ value at $A=4$ as $A^2$, as 
predicted by linear theory. Below that is shown the ratio of the energy to angular momentum of the outgoing GWs. The curves are fits of the form $a+bA^2$.
We indicate the numerical error by showing the results from the low and medium resolution runs
at $A=4$ and $12$.
Bottom: The changes in the BH quantities of total mass, angular momentum times the initial BH rotational frequency (upper panel), 
as well as irreducible mass, and dimensionless spin (lower panel) implied 
by the above fits.  We also show the changes in the BH 
that would occur if the outgoing GWs were always proportional to the ingoing GWs 
(labeled linear, i.e., ignoring the $A^2$ term of the fit shown in the top two panels).  
\label{egw_fig}
}
\end{figure}

The decreasing amplification with increasing wave amplitude is
also illustrated in Fig.~\ref{egw_fig}. We show how the total radiated energy $E_{\rm GW}$ scales with the 
ingoing amplitude.  After dividing out by the contribution
which goes as $A^2$, we find that the measured energy 
is well fit by including an additional $A^4$ term which decreases this quantity
at high amplitudes.  
We note that the convergence study indicates that the truncation error in the \emph{ratio} of 
quantities (e.g. $E_{\rm GW}$) from simulations with different $A$
is noticeably smaller than the absolute error in the quantity for either $A$
(suggesting that the leading-order error scales with the magnitude of the quantity).
We also show the ratio of $E_{\rm GW}$ to the radiated angular momentum
$J_{\rm GW}$.  For small amplitudes this ratio is very close to the value given by the dominant frequency of the 
ingoing GW, $\omega_0/m=0.375/M$, though for higher amplitudes this decreases,
with the trend also well fit by a quadratic expression in $A$.
In the bottom panel of Fig.~\ref{egw_fig} we show the estimated change in the BH parameters
using these extrapolations.  
For comparison, we also show the change that would occur 
if the energy and angular momentum absorbed by the BH followed
the linear scaling for all $A$. Assuming cosmic censorship, this
by itself implies the need for nonlinear corrections,
as otherwise it predicts $M_{\rm ir}$ should begin to decrease beyond $A \approx 22$.
Moreover, viewed as a quasistatic process, this behavior would begin to be problematic 
at even lower amplitudes ($A\approx 13$)
where $\Delta M_{\rm ir}$ reaches a maximum and starts to decrease. This occurs when 
the rotational frequency of the final BH 
(after losing energy to the GW) is $<\omega_0/m$.  
The fit to the simulation results, on the other hand, extends the range in $A$ where
$\Delta M_{\rm ir}$ shows a monotonic increase. This trend eventually reverses as well,
implying (again assuming cosmic censorship) that higher-order nonlinear terms 
would come into play.
Note that for $A\gtrsim 38$ the GW energy will exceed
that of the BH, and the problem becomes one of collapse of a GW
perturbed by a BH.

\begin{figure}
\begin{center}
\includegraphics[width=\columnwidth,draft=false]{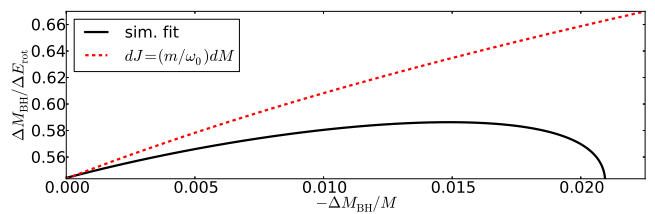}
\end{center}
\caption{
The efficiency of BH energy extraction 
$\eta$, as a
function of the energy extracted from the BH, that is implied by the fit 
to simulation results in Fig.~\ref{egw_fig} 
(lower amplitude branch).
We also show the efficiency expected for a process where the change in BH angular momentum 
to BH mass is given by $dJ=(m/\omega_0)dM$.
\label{eff_fig}
}
\end{figure}

In superradiant scattering, the BH loses some of its rotational energy $E_{\rm rot}=M_{\rm BH}-M_{\rm ir}$, part of which is carried away by the wave, and part of which
goes into increasing $M_{\rm ir}$, and hence is no longer extractable.  This process can
be viewed as having an efficiency $\eta=\Delta M_{\rm BH}/\Delta E_{\rm rot}$.
In Fig.~\ref{eff_fig} we show the efficiency obtained using the fits in Fig.~\ref{egw_fig}.  
The efficiency does initially increase with total extracted energy,
though it eventually reaches a maximum.  It is also always below the efficiency that would occur if the BH underwent a quasistatic process with $dJ= (m/\omega_0)dM$
(also shown), reminiscent of the fact that in thermodynamics, maximum efficiency 
is achieved only in a quasistatic process (e.g. a Carnot cycle).

In the extracted GWs we also see an indication of non-linear mode coupling, predominantly in 
the $l=|m|=4$ modes (here we refer to spin weight $-2$ spherical, not spheroidal, components),
which are absent in the ingoing GW (as verified by examining $\Psi_0$). These modes oscillate at twice the frequency of the $l=|m|=2$ modes.
The numerical errors on the higher modes are large (and indicate an underestimate), but the simulations indicate that the ratio
of the energy in the $(l,m)=(4,4)$ component to the $(2,2)$ component scales as $A^2$, with
$E_{44}/E_{22}\sim0.03$ at $A=12$.

\begin{figure}
\begin{center}
\includegraphics[width=\columnwidth,draft=false]{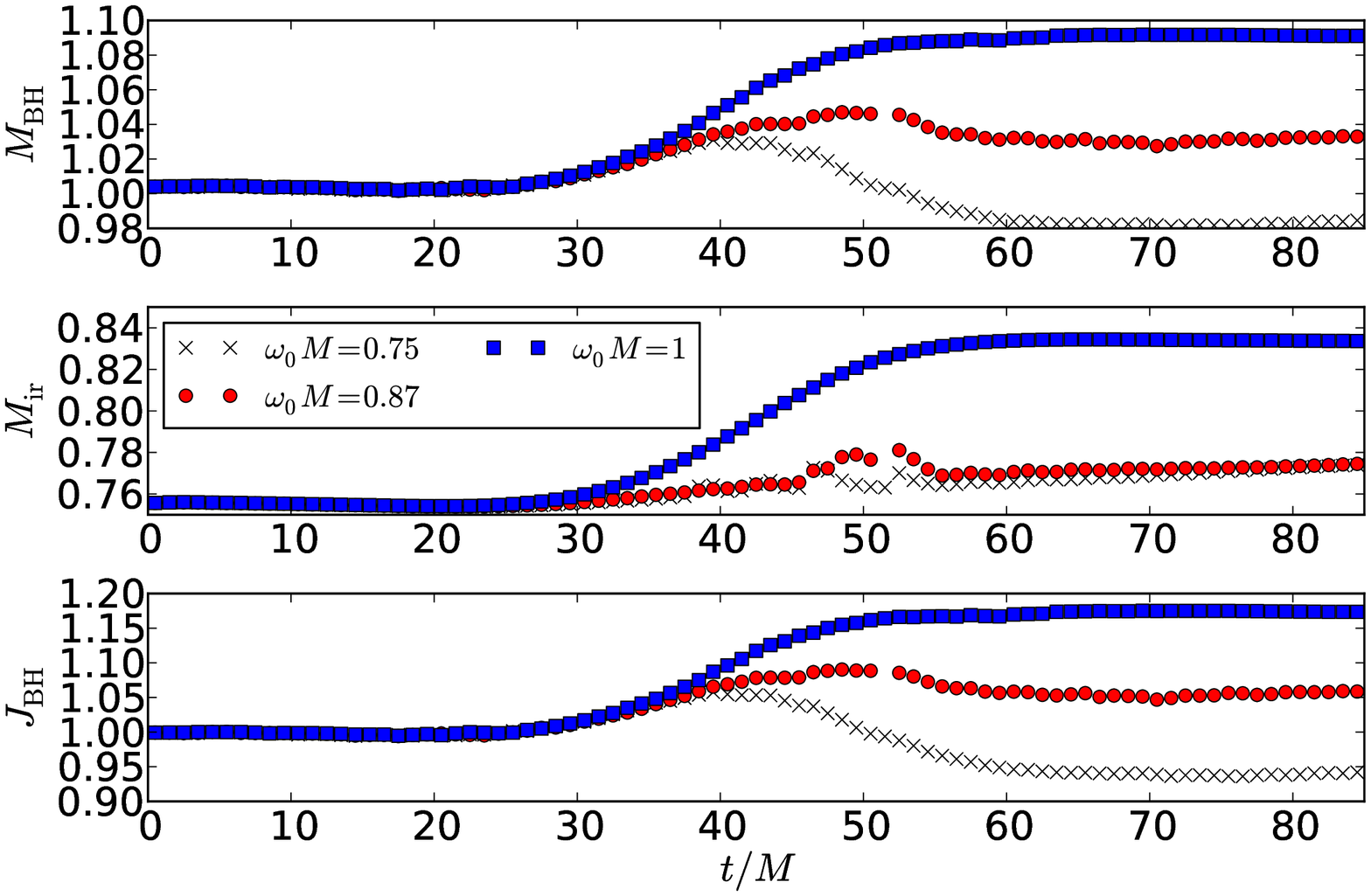}
\includegraphics[width=\columnwidth,draft=false]{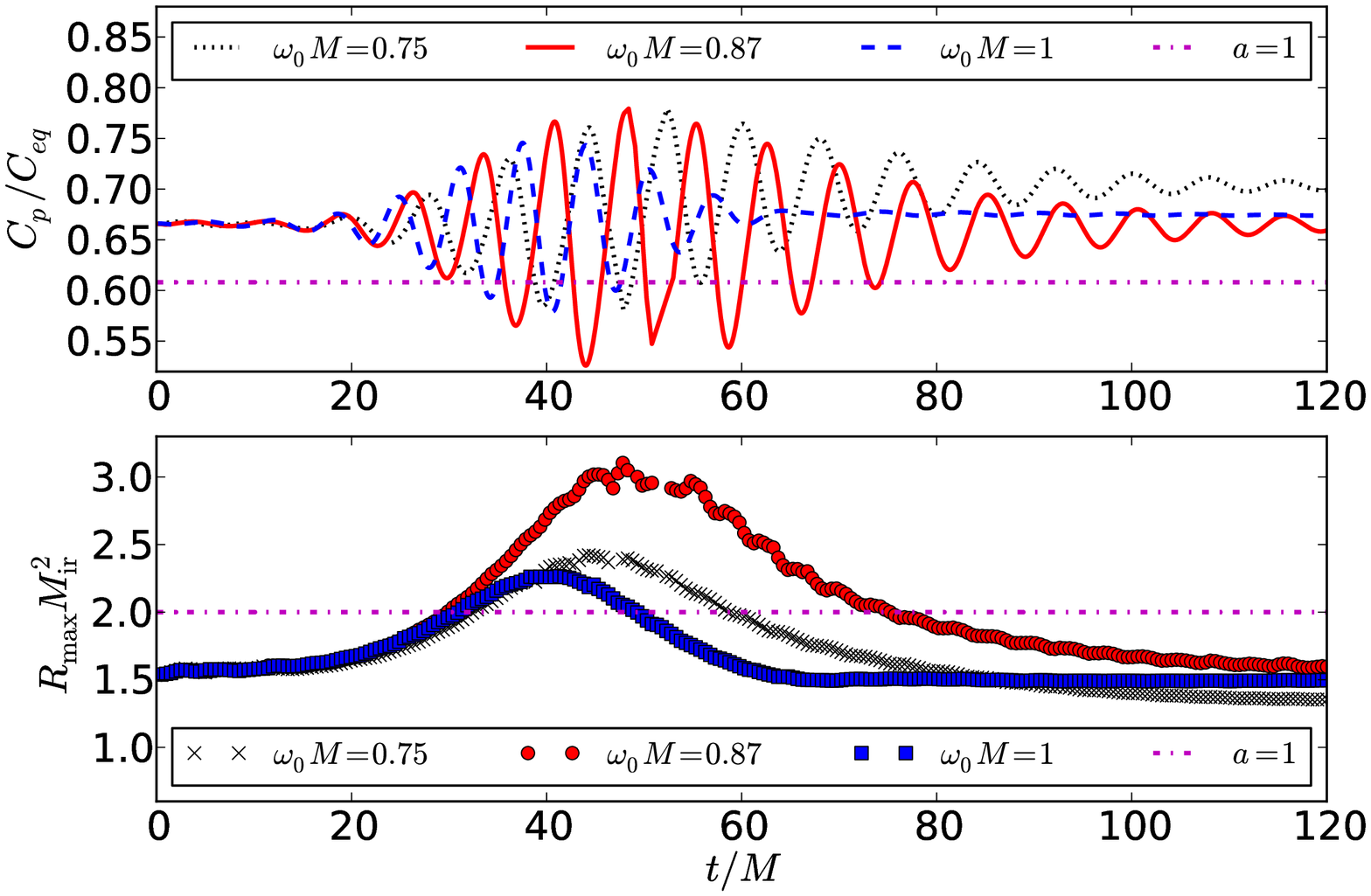}
\end{center}
\caption{
AH quantities during interaction with different frequency GW packets, each 
with initial mass $\approx 0.1M$.
Shown (in units where $M=1$) are the mass, irreducible mass, and angular momentum of the BH (top panel),
the ratio of proper polar to equatorial circumference (middle panel), and the maximum 
Ricci curvature of the horizon surface (bottom panel).
For the latter two quantities, we also show the values of an unperturbed BH with extremal spin.
\label{ah_fig}
}
\end{figure}

In Fig.~\ref{ah_fig} we show the behavior of the BH AH for the 
largest amplitude ($A=12$) case.  For comparison, we also show two other cases with the same mass
wave packet but with higher frequencies. 
In the $\omega_0 M=0.87$ case about $1/3$ of the ingoing GW energy is absorbed,
with the absorbed energy coming from the component
of the wave packet with frequency above  $m\Omega_{\rm BH}$ (see Fig.~\ref{eomega_fig}).  
For $\omega_0 M=1$, the GW is almost entirely absorbed by the BH. 
In contrast, the superradiance of the $\omega_0 M=0.75$ case is borne out by the
overall decrease in angular momentum $J_{\rm BH}$  and mass $M_{\rm BH}$ of the BH shown in the top panel.  
We note that during the highly dynamic
phase of the evolution, when the peak of the GW is incident on the BH there is no 
(even approximate) axisymmetric Killing vector, and hence no way of unambiguously defining $J_{\rm BH}$ 
(and therefore $M_{\rm BH}$, which we compute from $J_{\rm BH}$ and the irreducible 
mass $M_{\rm ir}$) in a coordinate-independent fashion. We simply calculate  
$J_{\rm BH}$ and $M_{\rm BH}$ using the axisymmetric Killing vector
of the isolated BH spacetime, and so 
some caution must be taken interpreting these quantities during the interaction.
 
Despite the opposite sign of the change in BH mass, the $\omega_0 M=0.75$ and 0.87 cases show a similar small increase in irreducible mass.
(We note that for these cases,  when the BH is highly distorted around $t\approx50M$, the AH finding algorithm is not always able to locate
the horizon to the desired tolerance, and the quantities computed around then are less accurate;
this accounts for the noisiness of these curves around that time.)
In the middle panel of Fig.~\ref{ah_fig} we show the ratio of the 
proper polar (as measured arbitrarily in the $y=0$ plane) and equatorial circumferences $C_{\rm p}/C_{\rm eq}$
of the AH indicating that the AH is becoming distorted from its axisymmetric shape and oscillating.  
The horizon distortions are further illustrated by the fact that in all three cases 
the maximum Ricci
scalar curvature of the AH 2-surface $R_{\rm AH}$ temporarily increases
above the value of an unperturbed, $a=1$ BH.  
The maximum value of $R_{\rm AH}$ occurs on the equator (and the minimum at the poles), 
as it does for an isolated spinning BH, though these AHs show large deviations from axisymmetry.

As can be seen in Fig.~\ref{ah_fig}, the largest distortions of the AH, in terms of oscillations of
$C_{\rm p}/C_{\rm eq}$ and $R_{\rm AH}$, occur not for the case that is overall superradiant, nor for the case
where the BH's mass and angular momentum change the most, 
but for $\omega_0 M =0.87$.  
This is presumably related to the fact that, for this case, the energy is peaked near the resonant frequency of the BH,
which approaches $m\Omega_{\rm BH}$ from above as $a\rightarrow 1$~\cite{Detweiler:1977gy}.

\section{Conclusions}
\label{conclusions}
In this paper, we presented results from full solutions of the Einstein equations describing 
the superradiant scattering of GW packets by a highly spinning BH. 
To our knowledge this is the first study of this phenomenon beyond the linear
and quasistatic regimes.
We demonstrated the reduction in the BH's rotational energy resulting
from this effect.
For large amplitude waves we found that backreaction 
serves to reduce the amplification of the scattered waves as well as reduce
the higher frequency content of the wave packet. The net result is 
a decrease in the efficiency of energy extraction, as expected for
the BH to avoid losing more energy than allowed by Hawking's area theorem.
Due to technical difficulties in following what becomes a highly dynamical
and distorted AH (a necessity for stable evolution with our excision method),
we were unable to study the interaction of waves with initial
energy $\gtrsim 0.1M$.
This would be an interesting area for future work, as
the extrapolation in Fig.~\ref{egw_fig} suggests then higher-order
effects are needed to prevent the BH from becoming superextremal.
We suspect that with the slicing employed here the intrinsic geometry
of the event horizon is well mimicked by that of the AH, 
though it would be interesting to 
explicitly study the event horizon
structure to confirm this.

We thank Amos Ori and Mihalis Dafermos for stimulating discussions that initiated this work.
This research was supported by NSF Grants No. PHY-1065710, No. PHY1305682, and the Simons Foundation (FP).
Simulations were run on the orbital cluster at Princeton University.

\bibliographystyle{h-physrev}
\bibliography{super}

\end{document}